\documentclass{moriond}

\usepackage{amsmath}
\usepackage{amssymb}

\newcommand{\D}{\mathrm{d}}

\def\be{\begin{equation}}
\def\ee{\end{equation}}
\def\bea{\begin{eqnarray}}
\def\eea{\end{eqnarray}}

\newcommand{\Photo}{}

\begin{document}
\vspace*{4cm}
\title{A holographic analysis of the pion}

\author{ Jeff Forshaw }

\address{Department of Physics and Astronomy,
  University of Manchester,
  Manchester M13 9PL,
  United Kingdom}

\author{ \underline{Ruben Sandapen} }

\address{Department of Physics, Acadia University, Wolfville, Nova Scotia, B4P 2R6, Canada}

\maketitle\abstracts{Being simultaneously a bound state and a pseudo-Goldstone boson of chiral symmetry breaking, the pion is an ideal probe of the intertwined phenomena of confinement and chiral symmetry breaking in QCD. Here, we compute the low energy pion observables with light-front holography, taking into account longitudinal dynamics using an ansatz that explores the strong degeneracy of the `t Hooft and Li-Vary models as well as their underlying connection in $\mathrm{AdS}_3$ as recently noted by Vegh.}

\section{Introduction}
In light-front QCD, the Schr\"odinger-like equation for the pion reads \cite{Brodsky:1997de} 
\begin{equation}
	\left(\frac{-\nabla^2}{x(1-x)} + \frac{m^2_q}{x(1-x)} +	U(x,\mathbf{b}) \right) \Psi (x,\mathbf{b}) = M^2 \Psi(x,\mathbf{b}), 
\label{LFSE-meson}
\end{equation}
where $M (m_q)$ is the meson (quark) mass and $\Psi(x,\mathbf{b})$ its light-front wavefunction. Here  $x$ is the light-front momentum fraction carried by the quark and $\mathbf{b}=(b_\perp, \varphi)$ is the transverse distance between the quark and antiquark.  The  normalization condition on the wavefunction is
\begin{equation}
	\int \mathrm{d}^2 \mathbf{b} \, \mathrm{d} x \, |\Psi(x,\mathbf{b})|^2 = 1 .
\label{normalization-b}
\end{equation}
Assuming that $U=U_\perp + U_\parallel$ and 
that the wavefunction factorizes as
\begin{equation}
	\Psi (x, \mathbf{b})= \frac{\phi (\zeta)}{\sqrt{2\pi \zeta}} e^{i L \varphi} X(x),\label{full-mesonwf}
\end{equation}
where $\boldsymbol{\zeta}=\sqrt{x(1-x)} \mathbf{b}$, Eq. \eqref{LFSE-meson} separates into
\begin{equation}
	\left[-\frac{\D^2}{\D \zeta^2}+\frac{4L^2-1}{4 \zeta^2}+U_\perp \right] \phi(\zeta)= M_\perp^2 \phi(\zeta)
\label{hSE}
\end{equation}
and
\begin{equation}
	 \left[\frac{m^2_q}{x(1-x)} +U_\parallel \right] X(x) =M_\parallel^2 X(x) 
\label{parallel}
\end{equation}
with $M^2=M_\perp^2 + M_\parallel^2$ and
\begin{equation}
	\int \mathrm{d} \zeta |\phi(\zeta)|^2 = \int \mathrm{d} x \frac{|X(x)|^2}{x(1-x)} =1 \;.
\end{equation}

Light-front holography uniquely fixes \cite{Brodsky:2013ar}
\begin{equation}
	U_\perp = U_{\mathrm{LFH}}=\kappa^4 \zeta^2 + 2 \kappa^2 (J-1) \;,
	\label{holographic-potential}
\end{equation}
where $J$ is spin of the meson and $\kappa$ is an emerging mass scale. With $\zeta=z_5$, where $z_5$ is the extra dimension in $\mathrm{AdS}_5$, Eq. \eqref{hSE} becomes the wave equation for freely propagating spin-$J$ string modes in  an $\mathrm{AdS}_5$ geometry warped by a quadratic dilaton, $\varphi(z_5) =e^{-\kappa^2 \zeta_5^2}$. Eq. \eqref{hSE} then predicts that \cite{Brodsky:2014yha} \begin{equation}
	M_{\perp}^2(n_\perp, J, L)= 4\kappa^2 \left( n_\perp + \frac{J+L}{2} \right) 
\label{MT}
\end{equation}
which tells us that the lowest lying state with $n=L=J=0$, i.e. the pion, is massless. This means that the pion mass is entirely generated by longitudinal dynamics: $M_\pi=M_\parallel$. On the other hand, the lowest lying eigenfunction of Eq. \eqref{hSE} is given by 
\begin{equation}
	\phi_{\pi}(\zeta) =\kappa \sqrt{2\zeta} \exp{\left(\frac{-\kappa^2 \zeta^2}{2}\right)} \;.
	\label{pionwf-gaussian}
\end{equation}


\section{Longitudinal dynamics}
\label{sec:Long}
Introducing $\chi(x)=X(x)/\sqrt{x(1-x)}$, Eq. \eqref{parallel} becomes
\begin{equation}
	 \left[\frac{m_q^2}{x(1-x)} + V_\parallel \right] \chi(x) =M_\parallel^2 \chi(x) 
\label{parallel-chi}
\end{equation}
where
\begin{equation}
  V_\parallel=  \frac{1}{\sqrt{x(1-x)}} U_\parallel \sqrt{x(1-x)} 
\label{V-U-parallel}
\end{equation}
and 
\begin{equation}
\int_0^1 \mathrm{d} x \, |\chi(x)|^2=1 \;. 
\label{normalization-chi}
\end{equation}

As first noted in \cite{Chabysheva:2012fe}, Eq. \eqref{parallel-chi} coincides with the 't Hooft (tH) Equation of large $N_c$ $\mathrm{QCD}_2$ if  
\begin{equation}
	V_\parallel= g^2 \mathcal{P} \int_0^1 {\rm d}y \frac{\chi(y)-\chi(x)}{(x-y)^2}
\end{equation}
where $\mathcal{P}$ denotes a Principal Value prescription \cite{tHooft:1974pnl}. An alternative model for $V_\parallel$ is that of Li and Vary (LV) \cite{Li:2015zda,Li:2021jqb}
\begin{equation}
	V_\parallel = -\sigma^2 \partial_x (x(1-x) \partial_x) \;.
\label{LV}
\end{equation}
Despite their explicitly different analytical forms, it is known \cite{Weller:2021wog} that the LV and tH potentials are degenerate in describing the pion data, with a longitudinal mode $\chi(x) \propto (x(1-x))^\beta$ where $\beta=m_q/\sigma$ and $\beta \propto m_q/g$ respectively.

Here, we propose the FS model:
\begin{equation}
\left(\frac{m^2_q-g^2}{x(1-x)}\right)\chi(x) - g^2 \mathcal{P} \int_0^1 {\rm d}y \frac{\chi(y)}{(x-y)^2}-\sigma^2 \partial_x (x(1-x) \partial_x) \chi(x)=M_\parallel^2 \chi(x) .
  \label{FSA}
\end{equation}
If $g=0$, we recover the LV Equation and if $\sigma=0$, we recover the tH Equation. If instead we choose  $m_q^2=g^2+\sigma^2/4$, Eq. \eqref{FSA} becomes 
\begin{equation}
-\mathcal{P}\int_0^1 {\rm d}z^\prime \frac{\chi(z^\prime)}{(z-z^\prime)^2}-\frac{1}{4g_s}  \sqrt{z(1-z)} \partial^2_z (\sqrt{z(1-z)} \chi(z))=\mu^2  \chi(z) \;,
	\label{AdS2}
\end{equation}
which is an $\mathrm{AdS}_3$ string equation derived by Vegh: \cite{Vegh:2023snc} 
\begin{equation}
-\mathcal{P}\int_0^1 {\rm d}z^\prime \frac{\chi(z^\prime)}{(z-z^\prime)^2}-\frac{1}{4g_s}\partial_z (z(1-z) \partial_z) \chi(z)=\mu^2  \chi(z) \;,
	\label{AdS1}
\end{equation}
where $g_s$ is the string tension in units of the AdS radius (squared). 

Note that under the transformation 
\begin{equation}
	V_\parallel \to h(x) V_\parallel \frac{1}{h(x)} 
\end{equation}
the eigenspectrum of Eq. \eqref{parallel-chi} is invariant while the eigenfunctions transform as $\chi(x) \to h(x) \chi(x)$. Motivated by the power law ground state solutions to the tH and LV equations, we consider $h(x) = [x(1-x)]^{n/2}$ and take $V_\parallel$ to be the tH or LV operator as our default (i.e. $n=0$). It turns out that $n<0$ is badly behaved in the chiral limit and that the pion data exclude $n\ge3$ while the remaining possibilities, $n=0,1,2$ are all able to fit the data \cite{Forshaw:2024mrh}. We refer to the $n=0,1,2$ cases as the FS-A, FS-B and FS-C models respectively.

\section{Pion observables}
The pion mass is given by Eq. \eqref{parallel-chi}: 
\begin{equation}
	M_\pi^2 = \int_0^1 \mathrm{d}x \; \chi^*(x) \left[\frac{m_q^2}{x(1-x)} + V_\parallel \right] \chi(x) 
	\label{Mpi-chi} 
\end{equation}
while the pion decay constant, EM form factor and pion-to-photon transition form factor are given by \cite{Forshaw:2024mrh}
\begin{align}
    f_\pi &= \sqrt{\frac{6}{\pi}} \int_0^1 \mathrm{d} x \;\Psi_\pi(x,\mathbf{0}) \nonumber \\ &= \frac{\sqrt{6}}{\pi} \kappa \int_0^1 \mathrm {d} x \sqrt{x(1-x)} \chi(x) \;,
\label{fpi}
\end{align}
\begin{align}
    F_\pi(Q^2) & = \pi \int_0^1 \mathrm{d} x \; \D b_\perp^2 \; J_0((1-x) b_\perp Q) \; |\Psi_\pi(x,\mathbf{b})|^2 \nonumber \\ & = \int_0^1 \mathrm{d} x \; |\chi(x)|^2 \; \exp \left(-\frac{(1-x)}{x} \frac{Q^2}{4\kappa^2} \right) 
\label{EMFF}
\end{align}
and
\begin{align}
\label{TFF}
	Q^2 F_{\pi \gamma} (Q^2)&=\frac{2 \kappa}{\sqrt{3} \pi}  \int_0^1 \mathrm{d} x \sqrt{x(1-x)} \chi(x) \\ \nonumber 
	\times & \int_0^\infty \mathrm{d} b_\perp (m_q b_\perp) K_1(m_q b_\perp) 
	 \exp\left(-\frac{\kappa^2 x(1-x) b_\perp^2}{2} \right) 
	Q J_1(b_\perp (1-x) Q) 
\end{align}
respectively. The pion charge radius is related to the $Q^2 \to 0$ limit of the form factor :
\begin{align}
    r^2_{\pi}&=  -6 \lim_{Q^2 \to 0}\frac{\mathrm{d}F_{\pi}(Q^2)}{\mathrm{d}Q^2} \nonumber \\ 
     & =  \frac{3}{2\kappa^2} \int_0^1 \mathrm{d} x \frac{(1-x)}{x} |\chi(x)|^2,
\label{rpi}
\end{align}
while the $\pi^0 \to \gamma \gamma$ decay width is related to the $Q^2 \to 0$ limit of the transition form factor : 
\begin{equation}
	\Gamma_{\gamma \gamma} = \frac{\pi}{4} \alpha_{\mathrm{em}}^2 M_\pi^3 |F_{\pi \gamma}(0)|^2 \; .
    \label{eq:gwidth}
\end{equation}

\section{Results}
  
\begin{figure}[htbp]
\centering 
\includegraphics[width=8cm]{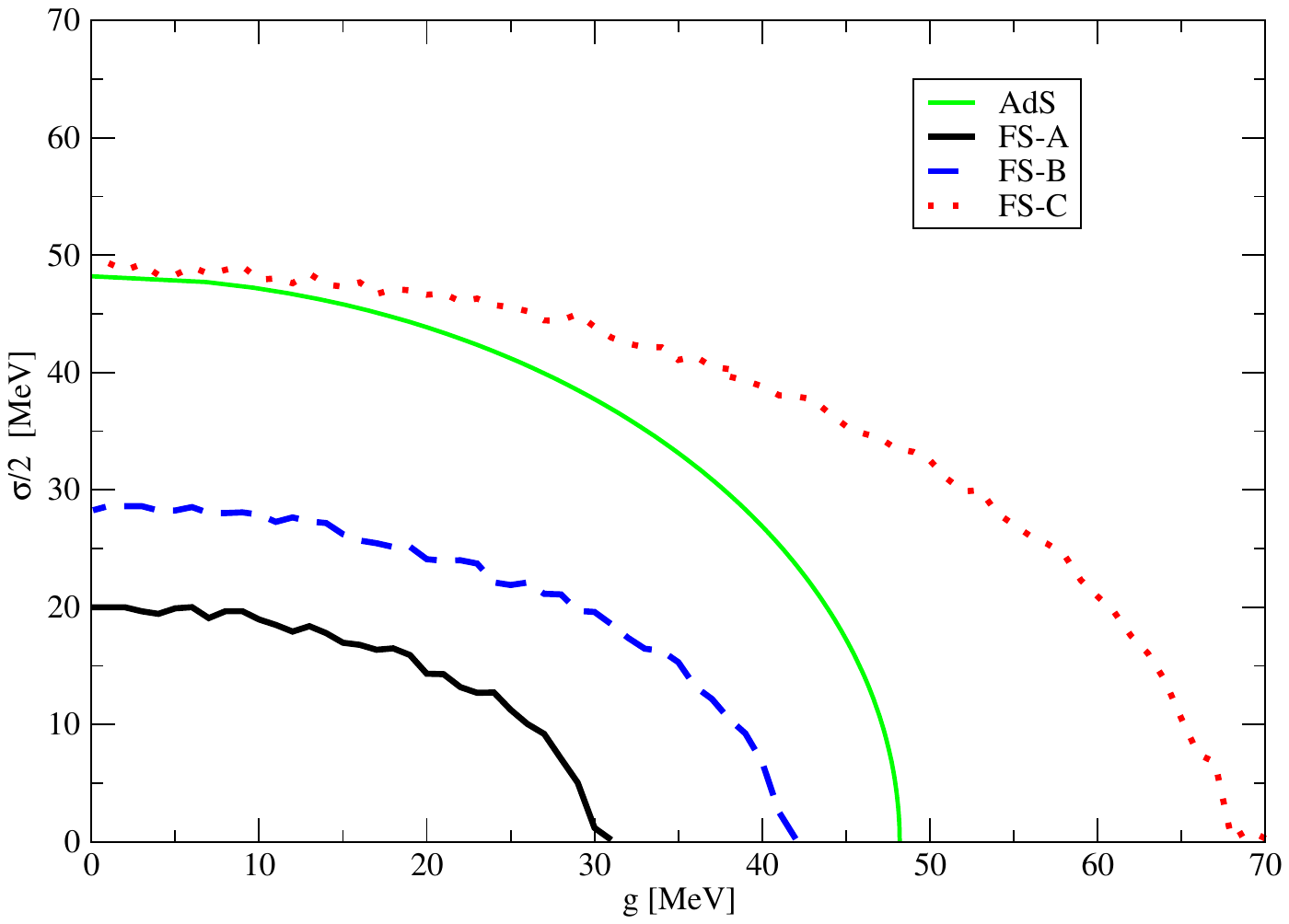} 
\caption{Exploring the degeneracy between the tH and LV models. Note that $m_q$ varies very weakly along each curve and that $\kappa=423$ MeV. }
\label{Fig:contours}
\end{figure}

In Fig.~\ref{Fig:contours}, we use the FS models to explore the degeneracy between the tH and LV models. Each point on a given curve represents a pair $(g,\sigma)$ values that accommodates the low energy pion data. Interestingly, the FS-C model is able to accommodate the holographic constraint, $g^2 + \sigma^2/4=m_q^2$ (green curve) in  the $\sigma \gg g$ limit, which is equivalent to a weak string coupling, $g_s \ll 1$.  


\section*{Acknowledgments}
RS thanks the organizers of the $59^{\mathrm{th}}$ Rencontres de Moriond for a very enjoyable conference. This research is supported by the Harrison McCain Foundation at Acadia University and the Natural Sciences and Engineering Research Council of Canada (NSERC).

\section*{References}

\bibliography{ref.bib}

\end{document}